\documentclass{elsart}

\usepackage{graphicx}

\usepackage{amssymb}
\newcommand{\ac}[1]{c:\Sigma\times\Sigma^{\leq{#1}}\rightarrow\Delta^{+}}
\newcommand{\eac}{\overline{c}:\Sigma^{*}\rightarrow\Delta^{*}}
\newcommand{\sstring}[1]{\sigma_{1}\sigma_{2}\ldots\sigma_{#1}}

\begin{document}

\begin{frontmatter}



\title{High Performance BWT-based Encoders
	\thanksref{CNCSIS}}
\thanks[CNCSIS]{Research supported in part by CNCSIS grant 632/2004.}


\author{Drago\c s N. Trinc\u a}
\address{Faculty of Computer Science, ``Al.I.Cuza'' University, 700483 Iasi, Romania}
\author{}\vspace*{-1.55\baselineskip}
\ead{dragost@infoiasi.ro}

\begin{abstract}
	In 1994, Burrows and Wheeler \cite{bw1} developed a data compression algorithm
	which performs significantly better than Lempel-Ziv based algorithms.
	Since then, a lot of work has been done in order to improve their algorithm, which is based on
	a reversible transformation of the input string, called BWT (the Burrows-Wheeler transformation).
	In this paper, we propose a compression scheme based on BWT, MTF (move-to-front coding),
	and a version of the algorithms presented in \cite{t2}.
\end{abstract}


\end{frontmatter}

\section{Introduction}
	A very promising development in the field of lossless data compression is the algorithm by Burrows and
	Wheeler \cite{bw1}.
	Since its publication in 1994, their algorithm has been widely studied, improved, and implemented on
	different platforms. Their original algorithm, as reported in \cite{bw1}, achieves speed comparable
	to Lempel-Ziv based algorithms and compression performance close to the best PPM techniques
	\cite{bwc1}.

	The most interesting and unusual step in their compression scheme is a reversible transformation of the
	input string (the Burrows-Wheeler transformation, or BWT), which reorders the symbols such that
	the newly created string contains the same symbols,
	but is easier to compress with simple locally adaptive algorithms such as move-to-front coding (MTF)
	\cite{bstw1}.

	In this paper, we propose a compression scheme based on BWT, MTF, adaptive codes \cite{t2,t3},
	and a version of the algorithms presented in \cite{t2}.
	More specifically, the following sections are aimed to present a detailed description
	of our algorithm in a progressive manner, including reports of experimental results.
	As we shall see, experiments performed on various well-known proteins prove that on this type of information
	our algorithm significantly outperforms the bzip2 utility \cite{js1},
	which is a well-known implementation of the algorithm introduced by Burrows and Wheeler.

\section{Adaptive codes}
	Adaptive codes have been recently presented in \cite {t2,t3}
	as a new class of non-standard variable-length codes.
	The aim of this section is to briefly review some basic definitions and notations. For more details,
	the reader is referred to \cite{t2,t3}.
	
	We denote by $|S|$ the \textit{cardinality} of the set $S$; if $x$ is a string of
	finite length, then $|x|$ denotes the length of $x$.
	The \textit{empty string} is denoted by $\lambda$.
	For an alphabet $\Delta$,
	we denote by $\Delta^{n}$ the set
	$\{s_{1}s_{2}\ldots{s_{n}} \mid s_{i}\in\Delta$ for all $i\}$,
	by $\Delta^{*}$ the set
	$\bigcup_{n=0}^{\infty}\Delta^{n}$,
	and by $\Delta^{+}$ the set
	$\bigcup_{n=1}^{\infty}\Delta^{n}$,
	where $\Delta^{0}$ denotes the set $\{\lambda\}$.
	Also, we denote by
	$\Delta^{\leq n}$ the set
	$\bigcup_{i=0}^{n}\Delta^{i}$,
	and by $\Delta^{\geq n}$ the set
	$\bigcup_{i=n}^{\infty}\Delta^{i}$.

	Let $X$ be a finite and nonempty subset of $\Delta^{+}$, and $w\in\Delta^{+}$.
	A \textit{decomposition of w} over $X$ is any sequence of strings
	$u_{1}, u_{2}, \ldots, u_{h}$ with $u_{i}\in X$, $1\leq i\leq h$, 
	such that $w=u_{1}u_{2}\ldots u_{h}$.
	A \textit{code} over $\Delta$ is any nonempty set $C\subseteq\Delta^{+}$ such
	that each string $w\in\Delta^{+}$ has at most one decomposition over $C$.
	A \textit{prefix code} over $\Delta$ is any code $C$ over $\Delta$ such that
	no string in $C$ is proper prefix of another string in $C$.
	If $u,v$ are two strings, then we denote by $u\cdot{v}$, or simply by $uv$ the catenation of
	$u$ with $v$.
\begin{defn}
	Let $\Sigma$ and $\Delta$ be alphabets. A function
	$\ac{n}$, $n\geq{1}$, is called \textup{adaptive code of order $n$} if its unique
	homomorphic extension $\eac$ defined by:
\begin{itemize}
\item $\overline{c}(\lambda)=\lambda$ 
\item $\overline{c}(\sstring{m})=$
	$c(\sigma_{1},\lambda)$
	$c(\sigma_{2},\sigma_{1})$
	$\ldots$
	$c(\sigma_{n-1},\sstring{n-2})$
	\newline
	$c(\sigma_{n},\sstring{n-1})$
	$c(\sigma_{n+1},\sstring{n})$
	$c(\sigma_{n+2},\sigma_{2}\sigma_{3}\ldots\sigma_{n+1})$
	\newline
	$c(\sigma_{n+3},\sigma_{3}\sigma_{4}\ldots\sigma_{n+2})\ldots$
	$c(\sigma_{m},\sigma_{m-n}\sigma_{m-n+1}\ldots\sigma_{m-1})$
\end{itemize}
	for all $\sstring{m}\in\Sigma^{+}$, is injective.
\end{defn}
	As it is clearly specified in the definition above, an adaptive code of order $n$ associates
	a variable-length codeword to the symbol being encoded depending on the previous $n$ symbols
	in the input data string.
	Let us take an example in order to better understand this mechanism.
\begin{exmp}
	Let $\Sigma=\{\texttt{\textup{a}},\texttt{\textup{b}},\texttt{\textup{c}}\}$, $\Delta=\{0,1\}$
	be alphabets, and
	$\ac{2}$ a function constructed by the following table.
	One can easily verify that $\overline{c}$ is injective, and according to Definition 1, $c$ is
	an adaptive code of order two.
\begin{table}[h]
\begin{center}
\textup{
Table 1. An adaptive code of order two.
\begin{tabular}{|c|c|c|c|c|c|c|c|c|c|c|c|c|c|}
\hline
$\Sigma\backslash\Sigma^{\leq{2}}$
		& $\texttt{\textup{a}}$  & $\texttt{\textup{b}}$  & $\texttt{\textup{c}}$  & $\texttt{\textup{aa}}$ &
		$\texttt{\textup{ab}}$ & $\texttt{\textup{ac}}$ & $\texttt{\textup{ba}}$ & $\texttt{\textup{bb}}$ &
		$\texttt{\textup{bc}}$ & $\texttt{\textup{ca}}$ & $\texttt{\textup{cb}}$ & $\texttt{\textup{cc}}$ &
		$\lambda$
																	\\ 	\hline
$\texttt{\textup{a}}$ & $01$ & $10$ & $10$ & $00$ & $11$ & $10$ & $01$ & $10$ & $11$ & $11$ & $11$ & $0$ & $0$
																	\\	\hline
$\texttt{\textup{b}}$ & $10$ & $00$ & $11$ & $11$ & $01$ & $00$ & $00$ & $11$ & $01$ & $10$ & $00$ & $10$ & $11$
																	\\	\hline
$\texttt{\textup{c}}$ & $11$ & $01$ & $01$ & $10$ & $00$ & $11$ & $11$ & $00$ & $00$ & $00$ & $10$ & $11$ & $10$
																	\\	\hline
\end{tabular}}
\end{center}
\end{table}
\newline
	Let $x=\texttt{\textup{abacca}}\in\Sigma^{+}$ be an input data string. 
	Using the definition above, we encode $x$ by
	$\overline{c}(x)=
	c(\texttt{\textup{a}},\lambda)c(\texttt{\textup{b}},\texttt{\textup{a}})
	c(\texttt{\textup{a}},\texttt{\textup{ab}})c(\texttt{\textup{c}},\texttt{\textup{ba}})
	c(\texttt{\textup{c}},\texttt{\textup{ac}})c(\texttt{\textup{a}},\texttt{\textup{cc}})=
	0101111110$.
\end{exmp}
	Let $\ac{n}$ be an adaptive code of order $n$, $n\geq{1}$. We denote by
	$C_{c, \sigma_{1}\sigma_{2}\ldots\sigma_{h}}$ the set
	$\{c(\sigma,\sigma_{1}\sigma_{2}\ldots\sigma_{h}) \mid \sigma\in\Sigma\}$,
	for all $\sigma_{1}\sigma_{2}\ldots\sigma_{h}\in\Sigma^{\leq{n}}-\{\lambda\}$,
	and by $C_{c, \lambda}$ the set $\{c(\sigma,\lambda) \mid \sigma\in\Sigma\}$.
	We write $C_{\sigma_{1}\sigma_{2}\ldots\sigma_{h}}$ instead of
	$C_{c, \sigma_{1}\sigma_{2}\ldots\sigma_{h}}$,
	and $C_{\lambda}$ instead of $C_{c, \lambda}$
	whenever there is no confusion.
	Let us denote by $AC(\Sigma,\Delta,n)$ the set
	$\{\ac{n} \mid  c$ is an adaptive code of order $n\}$.
\begin{thm}
	Let $\Sigma$ and $\Delta$ be two alphabets and $\ac{n}$ a function, $n\geq{1}$.
	If $C_{u}$ is prefix code, for all $u\in\Sigma^{\leq{n}}$, then $c\in{AC(\Sigma,\Delta,n)}$.
\end{thm}

\section{A high performance BWT-based compression scheme}
	As we have already pointed out, the algorithm introduced in 1994 by Burrows and Wheeler
	\cite{bw1} is one of the greatest developments in the field of lossless
	data compression.

	Their algorithm has received special attention not only for its Lempel-Ziv like execution
	speed and compression performance close to the best statistical modelling techniques \cite{bwc1},
	but also for the
	algorithms it combines. Let us briefly describe the three steps in their compression scheme.
\begin{description}
\item[BWT.] Let $S$ be a string of length $n$ which is to be compressed.
	The idea is to apply a reversible transformation (called BWT, or the Burrows-Wheeler transformation)
	to the string $S$ in order to form a new string $S'$,
	which contains the same symbols. The purpose of this transformation is to group together instances
	of a symbol $x_{i}$ occurring in $S$.
	More precisely, if a symbol $x_{i}$ is very often followed by $x_{j}$ in $S$, then the occurrences
	of $x_{i}$ tend to be grouped together in $S'$. Thus, $S'$ has a high locality of reference and
	is easier to compress with simple locally adaptive compression schemes such as move-to-front coding
	(MTF). 
\item[MTF.] The idea of move-to-front coding (MTF) is based on self-organizing linear lists.
	Let $L$ be a linear list containing the symbols which occur in $S'$.
	If $x_{i}$ is the current symbol in $S'$ which is to be encoded, then the encoder looks up the
	current position of $x_{i}$ in $L$, outputs that position and updates $L$ by moving $x_{i}$
	to the front of the list.
\item[EC.]
	A final entropy coding (EC) step follows the move-to-front encoder.
	Since the output of MTF usually consists of small integers, it can be efficiently encoded
	using a Huffman encoder.
\end{description}
	This is the algorithm which has led to the development of one of the best techniques in
	the field of lossless data compression.
	Let us present a detailed description of BWT and MTF,
	since our encoder is also based on these algorithms.
\\\\
{\bf Algorithm BWT.}
{\em
	Let $\Sigma=\{\sigma_{0},\sigma_{1},\ldots,\sigma_{p-1}\}$ be an ordered set,
	and let $S=s_{1}s_{2}\ldots{s_{n}}$ be a string
	over $\Sigma$, that is, $s_{i}\in\Sigma$ for all $i\in\{1,2,\ldots,n\}$.
	If $M$ is a matrix, $M[i,j]$ denotes the $j$-th element (from left to right)
	of the $i$-th row.
\textup{
\\\\
{\bf INPUT:} the string $S=s_{1}s_{2}\ldots{s_{n}}$ of length $n$.
\begin{enumerate}
\item[1.  ]
	Let $M$ be a $n\times{n}$ matrix whose elements are symbols, and whose rows
	are the rotations (cyclic shifts)
	of $S$, sorted in lexicographical order.
	Precisely, if $s_{k_{1}}s_{k_{2}}\ldots{s_{k_{n}}}$ is the $i$-th rotation of $S$
	(in lexicographical order), then $M[i,j]=s_{k_{j}}$ for all $j\in\{1,2,\ldots,n\}$.
\item[2.  ]
	Let $I$ be the index of the first row in $M$ which
	contains the string $S$ (there is at least one such row).
	Exactly, $I$ is the smallest integer such that $M[I,j]=s_{j}$ for all $j\in\{1,2,\ldots,n\}$.
\item[3.  ]
	Let $S'=t_{1}t_{2}\ldots{t_{n}}$ be the string contained in the last column of the matrix $M$,
	that is, $t_{i}=M[i,n]$ for all $i\in\{1,2,\ldots,n\}$.
\end{enumerate}
{\bf OUTPUT:} the $2$-tuple $(S',I)$.\\\\
}
	Interestingly enough, there exists an efficient algorithm which reconstructs the original string $S$
	using only $S'$ and $I$. However, the paper by Burrows and Wheeler \cite{bw1} gives a very detailed
	description of this algorithm, so we won't get into it.
	Instead, let us explain why the transformed string $S'$ compresses much better than $S$.
	Consider a symbol $x_{i}$ which is very often followed by $x_{j}$ in $S$. Since the rows of $M$ are
	the sorted rotations of $S$, and the symbol $M[i,n]$ precedes the symbol $M[i,1]$ in $S$, for all
	$i\in\{1,2,\ldots,n\}$, some consecutive rotations that start with $x_{j}$ are likely to end in $x_{i}$.
	This is why $S'$ has a high locality of reference, and is easier to compress with locally adaptive
	compression schemes such as MTF.
}
\\\\
	Let us now introduce some useful notation.
	Let $\mathcal{U}=(u_{1},u_{2},\ldots,u_{k})$ be a $k$-tuple.
	We denote by $\mathcal{U}.i$ the $i$-th component of $\mathcal{U}$,
	that is, $\mathcal{U}.i=u_{i}$ for all $i\in\{1,2,\ldots,k\}$.
	The $0$-tuple is denoted by $()$. The length of a tuple $\mathcal{U}$ is denoted by $Len(\mathcal{U})$.
	If $\mathcal{V}=(v_{1},v_{2},\ldots,v_{b})$,
	$\mathcal{M}=(m_{1},m_{2},\ldots,m_{r},\mathcal{U})$, $\mathcal{N}=(n_{1},n_{2},\ldots,n_{s},\mathcal{V})$,
	$\mathcal{P}=(p_{1},\ldots,p_{i-1},p_{i},p_{i+1},\ldots,p_{t})$ are tuples, and q is an element or a tuple,
	then we define
	$\mathcal{P}\vartriangleleft{q}$, $\mathcal{P}\vartriangleright{i}$,
	$\mathcal{U}\vartriangle{\mathcal{V}}$, and $\mathcal{M}\lozenge{\mathcal{N}}$ by:
\begin{itemize}
\item $\mathcal{P}\vartriangleleft{q}=(p_{1},\ldots,p_{t},q)$
\item $\mathcal{P}\vartriangleright{i}=(p_{1},\ldots,p_{i-1},p_{i+1},\ldots,p_{t})$
\item $\mathcal{U}\vartriangle{\mathcal{V}}=(u_{1},u_{2},\ldots,u_{k},v_{1},v_{2},\ldots,v_{b})$
\item $\mathcal{M}\lozenge{\mathcal{N}}=$
	$(m_{1}+n_{1},m_{2}+1,\ldots,m_{r}+1,n_{2}+1,\ldots,n_{s}+1,\mathcal{U}\vartriangle{\mathcal{V}})$
\end{itemize}
	where $m_{1},m_{2},\ldots,m_{r},n_{1},n_{2},\ldots,n_{s}$ are integers.
\\\\
{\bf Algorithm MTF.}
{\em
	Let $S'=t_{1}t_{2}\ldots{t_{n}}$ be the string obtained above.
	The MTF encoder works as follows.
\textup{
\\\\
{\bf INPUT:} the string $S'=t_{1}t_{2}\ldots{t_{n}}$ of length $n$.
\begin{enumerate}
\item[1.  ]
	Consider a linear list $L$ which contains the symbols occurring in $S'$ exactly once,
	sorted in lexicographical order. Also, let $\mathcal{R}=()$.
\item[2.  ]
	For each $i=1,2,\ldots,n$ execute:
\begin{enumerate}
\item[]
	2.1.\hspace{10pt}Let $q$ be the number of elements preceding $t_{i}$ in $L$.
\item[]
	2.2.\hspace{10pt}$\mathcal{R}:=\mathcal{R}\vartriangleleft{q}$.
\item[]
	2.3.\hspace{10pt}In the list $L$, move $t_{i}$ to the front of the list.
\end{enumerate}
\end{enumerate}
{\bf OUTPUT:} the $n$-tuple $\mathcal{R}$.
}
}
\begin{exmp}
	Let $\Sigma=\{\texttt{\textup{a}},\texttt{\textup{c}},\texttt{\textup{e}},\texttt{\textup{h}},
	\texttt{\textup{r}},\texttt{\textup{s}}\}$ be an alphabet, and consider the string
	$S=\texttt{\textup{research}}$ over $\Sigma$.
	One can verify that:
\begin{center}
	$M = 
	\left[
	\begin{array}{cccccccc}
	\texttt{\textup{a}}\hspace{8pt} & \texttt{\textup{r}}\hspace{8pt} & \texttt{\textup{c}}\hspace{8pt} &
	\texttt{\textup{h}}\hspace{8pt} & \texttt{\textup{r}}\hspace{8pt} & \texttt{\textup{e}}\hspace{8pt} &
	\texttt{\textup{s}}\hspace{8pt} & \texttt{\textup{e}} \\
	\texttt{\textup{c}}\hspace{8pt} & \texttt{\textup{h}}\hspace{8pt} & \texttt{\textup{r}}\hspace{8pt} &
	\texttt{\textup{e}}\hspace{8pt} & \texttt{\textup{s}}\hspace{8pt} & \texttt{\textup{e}}\hspace{8pt} &
	\texttt{\textup{a}}\hspace{8pt} & \texttt{\textup{r}} \\
	\texttt{\textup{e}}\hspace{8pt} & \texttt{\textup{a}}\hspace{8pt} & \texttt{\textup{r}}\hspace{8pt} &
	\texttt{\textup{c}}\hspace{8pt} & \texttt{\textup{h}}\hspace{8pt} & \texttt{\textup{r}}\hspace{8pt} &
	\texttt{\textup{e}}\hspace{8pt} & \texttt{\textup{s}} \\
	\texttt{\textup{e}}\hspace{8pt} & \texttt{\textup{s}}\hspace{8pt} & \texttt{\textup{e}}\hspace{8pt} &
	\texttt{\textup{a}}\hspace{8pt} & \texttt{\textup{r}}\hspace{8pt} & \texttt{\textup{c}}\hspace{8pt} &
	\texttt{\textup{h}}\hspace{8pt} & \texttt{\textup{r}} \\
	\texttt{\textup{h}}\hspace{8pt} & \texttt{\textup{r}}\hspace{8pt} & \texttt{\textup{e}}\hspace{8pt} &
	\texttt{\textup{s}}\hspace{8pt} & \texttt{\textup{e}}\hspace{8pt} & \texttt{\textup{a}}\hspace{8pt} &
	\texttt{\textup{r}}\hspace{8pt} & \texttt{\textup{c}} \\
	\texttt{\textup{r}}\hspace{8pt} & \texttt{\textup{c}}\hspace{8pt} & \texttt{\textup{h}}\hspace{8pt} &
	\texttt{\textup{r}}\hspace{8pt} & \texttt{\textup{e}}\hspace{8pt} & \texttt{\textup{s}}\hspace{8pt} &
	\texttt{\textup{e}}\hspace{8pt} & \texttt{\textup{a}} \\
	\texttt{\textup{{\bf r}}}\hspace{8pt} & \texttt{\textup{{\bf e}}}\hspace{8pt} &
	\texttt{\textup{{\bf s}}}\hspace{8pt} & \texttt{\textup{{\bf e}}}\hspace{8pt} &
	\texttt{\textup{{\bf a}}}\hspace{8pt} & \texttt{\textup{{\bf r}}}\hspace{8pt} &
	\texttt{\textup{{\bf c}}}\hspace{8pt} & \texttt{\textup{{\bf h}}} \\
	\texttt{\textup{s}}\hspace{8pt} & \texttt{\textup{e}}\hspace{8pt} & \texttt{\textup{a}}\hspace{8pt} &
	\texttt{\textup{r}}\hspace{8pt} & \texttt{\textup{c}}\hspace{8pt} & \texttt{\textup{h}}\hspace{8pt} &
	\texttt{\textup{r}}\hspace{8pt} & \texttt{\textup{e}}
	\end{array}
	\right]$
\end{center}
	is the matrix containing the sorted rotations of $S$, $I=7$, $S'=\texttt{\textup{ersrcahe}}$,
	and $\mathcal{R}=(2,4,5,1,4,4,5,5)$.
\end{exmp}
	At this point, it should be clear that applying BWT and MTF as described so far, the compression of
	the string $S$ is reduced to the compression of the tuple $\mathcal{R}$.
	Also, it is trivial to see that if $S$ is sufficiently large (at least several kilobytes),
	then the tuple $\mathcal{R}$ will consist mostly of large blocks of zeroes.
	For other details on these algorithms (including implementation details) the reader is referred to
	\cite{bw1}.

	New algorithms for data compression, based on adaptive codes of order one, have been recently
	presented in \cite{t2,t3}, where we have behaviorally shown that for a large class of input strings,
	our algorithms substantially outperform the well-known Lempel-Ziv compression technique \cite{zl1,zl2}.
	The final encoder in our compression scheme is based on the algorithms proposed in \cite{t2}.
	Before describing it in great detail, let us review the Huffman algorithm,
	since our encoder is based partly on this well-known compression technique.
	For further details on the Huffman algorithm, the reader is referred to \cite{ng1,ds1}.
\\\\
{\bf Algorithm Huffman.}
{\em
	As described below, the well-known Huffman algorithm takes as input a tuple
	$\mathcal{F}=(f_{1},f_{2},\ldots,f_{n})$  of frequencies, and returns a tuple
	$\mathcal{V}=(v_{1},v_{2},\ldots,v_{n})$
	of codewords, such that $v_{i}$ is the codeword corresponding to the symbol with the frequency $f_{i}$,
	for all $i\in\{1,2,\ldots,n\}$.
\textup{
\\\\
{\bf INPUT:} a tuple $\mathcal{F}=(f_{1},f_{2},\ldots,f_{n})$ of frequencies.
\begin{enumerate}
\item[1.  ]
	Consider the $n$-tuples $\mathcal{L}=((f_{1},0,(1)),(f_{2},0,(2)),\ldots,(f_{n},0,(n)))$
	and $\mathcal{V}=(\lambda,\lambda,\ldots,\lambda)$.
\item[2.  ]
	If $n=1$ then $\mathcal{V}.1:=0$.
\item[3.  ]
	While $Len(\mathcal{L})>1$ execute:
\begin{enumerate}
\item[]
	3.1.\hspace{10pt}Let $i<j$ be the smallest integers such that $\mathcal{L}.i.1$, $\mathcal{L}.j.1$ are the
\item[]
	\hspace{29pt}smallest elements of the set $\{\mathcal{L}.q.1 \mid q\in\{1,2,\ldots,Len(\mathcal{L})\}\}$.
\item[]
	3.2.\hspace{10pt}$F:=\{\mathcal{L}.i.Len(\mathcal{L}.i).r \mid
				r\in\{1,2,\ldots,Len(\mathcal{L}.i.Len(\mathcal{L}.i))\}\}$.
\item[]
	3.3.\hspace{10pt}$S:=\{\mathcal{L}.j.Len(\mathcal{L}.j).r \mid
				r\in\{1,2,\ldots,Len(\mathcal{L}.j.Len(\mathcal{L}.j))\}\}$.
\item[]
	3.4.\hspace{10pt}For each $x\in{F}$ execute $\mathcal{V}.x:=0\cdot{\mathcal{V}.x}$.
\item[]
	3.5.\hspace{10pt}For each $x\in{S}$ execute $\mathcal{V}.x:=1\cdot{\mathcal{V}.x}$.
\item[]
	3.6.\hspace{10pt}$\mathcal{U}:=\mathcal{L}.i\lozenge{\mathcal{L}.j};$
	$\mathcal{L}:=\mathcal{L}\vartriangleright{j};$
	$\mathcal{L}:=\mathcal{L}\vartriangleright{i};$
	$\mathcal{L}:=\mathcal{L}\vartriangleleft{\mathcal{U}}$.
\end{enumerate}
\end{enumerate}
{\bf OUTPUT:} the tuple $\mathcal{V}$.
}
}
\\\\
{\bf Algorithm AE (Adaptive Encoder).}
{\em
	The final encoding step in our compression scheme is based on the
	algorithms presented in \cite{t2}, that is, on adaptive codes of order one.
	As we have already discussed, the input of this final encoder is the output of MTF,
	that is, the tuple $\mathcal{R}$.
{\em
	Let $\Sigma=\{\sigma_{1},\sigma_{2},\ldots,\sigma_{p}\}$ be an alphabet, and let $x$ be
	a string over $\Sigma$. Let $q$ be the number of symbols occurring in $x$ (thus, $q\leq{p}$).
	Let us explain the main idea of our scheme. Consider that $u\in\Sigma^{n}$ is some
	substring of the input string $x$. Also, let us denote by $\textit{Follow}(u)$ the set
	of symbols that follow the substring $u$ in $x$.
	For each symbol $c\in\textit{Follow}(u)$, let us denote by $\textit{Freq}(c,u)$
	the frequency of the substring $uc$ in $x$. One can easily remark that $\textit{Follow}(u)$
	cannot contain more than $q$ symbols. Moreover, in the most cases,
	the number of symbols in $\textit{Follow}(u)$ is significantly smaller than $q$.
	Instead of applying the Huffman's algorithm to the $q$ symbols occurring in $x$,
	we apply it to the set $\{\textit{Freq}(c,u) \mid c\in\textit{Follow}(u)\}$, since this
	set has a smaller number of frequencies. If $\textit{code}(c,u)$ is the codeword
	associated to $\textit{Freq}(c,u)$, then we encode $c$ by $\textit{code}(c,u)$ if it is
	preceded by $u$. Thus, we get smaller codewords.

	This procedure is actually applied to every substring $u$ of length $n$ occurring in $x$.
	Thus, we associate to each symbol a set of codewords, and encode every symbol with one of the
	codewords in its set, depending on the previous $n$ symbols. 
}
\begin{figure}[h]
\begin{picture}(233,385)(-54,0)
\put(2,385){\line(0,-1){385}}
\put(280,385){\line(0,-1){385}}
\put(2,0){\line(1,0){278}}
\put(2,385){\line(1,0){278}}

\put(5,375){{\bf INPUT:} a string $x=x_{1}x_{2}\ldots{x_{t}}\in\Sigma^{+}$.}
\put(5,351){\texttt{1.}Let $a:\Sigma\times\Sigma^{n}\rightarrow\{0,1\}^{*}$,
	$b:\Sigma\times\Sigma^{n}\rightarrow\{0,1\}$,}
\put(5,339){\hspace{12pt}$c:\Sigma\times\Sigma^{n}\rightarrow\mathbb{N}$ be three functions$;$}
\put(5,327){\texttt{2.}Let $d:\{1,2,\ldots,p^{n}\}\rightarrow\Sigma^{n}$
	be a bijective function$;$}
\put(5,315){\texttt{3.}for each $(\sigma,u)\in\Sigma\times\Sigma^{n}$ do}
\put(5,303){\hspace{12pt}\texttt{3.1.}$a(\sigma,u)\leftarrow\lambda;$}
\put(5,291){\hspace{12pt}\texttt{3.2.}$b(\sigma,u)\leftarrow{0};$}
\put(5,279){\hspace{12pt}\texttt{3.3.}$c(\sigma,u)\leftarrow{0};$}
\put(5,267){\texttt{4.}for $i=n+1$ to $t$ do}
\put(5,255){\hspace{12pt}\texttt{4.1.}$b(x_{i},x_{i-n}\ldots{x_{i-1}})\leftarrow{1};$}
\put(5,243){\hspace{12pt}\texttt{4.2.}$c(x_{i},x_{i-n}\ldots{x_{i-1}})\leftarrow{c(x_{i},x_{i-n}\ldots{x_{i-1}})}$\hspace{0pt}$+1;$}
\put(5,231){\texttt{5.}for $j=1$ to $p^{n}$ do}
\put(5,219){\hspace{12pt}\texttt{5.1.}$\mathcal{S}\leftarrow{()};k\leftarrow{1};$}
\put(5,207){\hspace{12pt}\texttt{5.2.}for $i=1$ to $p$ do}
\put(5,195){\hspace{36pt}\texttt{5.2.1.}if $b(\sigma_{i},d(j))=1$ then}
\put(5,183){\hspace{72pt}\texttt{5.2.1.1.}$\mathcal{S}\leftarrow\mathcal{S}\vartriangleleft{c(\sigma_{i},d(j))};$}
\put(5,171){\hspace{12pt}\texttt{5.3.}if $Len(\mathcal{S})\geq{2}$ then}
\put(5,159){\hspace{36pt}\texttt{5.3.1.}$\mathcal{V}\leftarrow{\textup{Huffman}(\mathcal{S})};$}
\put(5,147){\hspace{12pt}\texttt{5.4.}for $i=1$ to $p$ do}
\put(5,135){\hspace{36pt}\texttt{5.4.1.}if $b(\sigma_{i},d(j))=1$ then}
\put(5,123){\hspace{72pt}\texttt{5.4.1.1.}$a(\sigma_{i},d(j))\leftarrow\mathcal{V}.k;$}
\put(5,111){\hspace{72pt}\texttt{5.4.1.2.}$k\leftarrow{k+1};$}
\put(5,99){\texttt{6.}$\mathcal{Y}\leftarrow{()};Z\leftarrow\lambda;$}
\put(5,87){\texttt{7.}for $i=1$ to $p$ do}
\put(5,75){\hspace{12pt}\texttt{7.1.}for $j=1$ to $p^{n}$ do}
\put(5,63){\hspace{36pt}\texttt{7.1.1.}if $a(\sigma_{i},d(j))\neq\lambda$ then}
\put(5,51){\hspace{72pt}\texttt{7.1.1.1.}$\mathcal{Y}\leftarrow{\mathcal{Y}}\vartriangleleft{a(\sigma_{i},d(j))};$}
\put(5,39){\texttt{8.}for $i=n+1$ to $t$ do}
\put(5,27){\hspace{12pt}\texttt{8.1.}$Z\leftarrow{Z\cdot{a(x_{i},x_{i-n}\ldots{x_{i-1}})}};$}

\put(5,3){{\bf OUTPUT:} the tuple $(x_{1}x_{2}\ldots{x_{n}},b,\mathcal{Y},Z)$.}
\end{picture}
\caption{EAH$n$.}
\end{figure}
{\em
	The complete algorithm is given above. Let us now explain what exactly the algorithm
	performs at each step. The first three steps are aimed to initialize the functions needed.
	Note that the function $d$ actually allows us to access the elements of $\Sigma^{n}$
	in a certain order. In the fourth step, $b(x_{i},x_{i-n}\ldots{x_{i-1}})$ is switched
	to $1$, since the substring $x_{i-n}\ldots{x_{i-1}x_{i}}$ occurs at least once in $x$,
	and the frequency of $x_{i-n}\ldots{x_{i-1}x_{i}}$ is incremented.
	In the fifth step, for every substring $d(j)$ of length $n$, we apply the Huffman's algorithm
	to the symbols following $d(j)$ in $x$. In the next two steps, $\mathcal{Y}$ is a tuple of
	codewords constructed as follows.
	If $c\in\Sigma$ and $u\in\Sigma^{n}$, then $a(c,u)$ is appended to $\mathcal{Y}$ if and
	only if $a(c,u)\neq\lambda$, that is, if $c\in\textit{Follow}(u)$ and
	$|\textit{Follow}(u)|\geq{2}$.
	Finally, in the last step, $Z$ denotes the compression of $x_{n+1}\ldots{x_{t}}$.
	
	So, the compression of the string $x$ is actually $Z$. The first three components of the output
	($x_{1}x_{2}\ldots{x_{n}}$,$b$, and $\mathcal{Y}$) are only needed when decoding $Z$ into $x$.
}
\\\\
	Let us now take an example in order to better understand the description above.
\begin{exmp}
	Let $\Sigma=\{\texttt{\textup{a}},\texttt{\textup{b}}\}$ be an alphabet,
	and let us take $x=\texttt{\textup{baabbabab}}\in\Sigma^{+}$ as
	an input data string. After applying \textup{EAH}$2$ to $x$, we get the results reported in the
	tables below.
\begin{table}[h]
\caption{The function $a$ after EAH$2(x)$.}
\begin{center}
\textup{
\begin{tabular}{|c|c|c|c|c|c|c|c|c|c|}
\hline
\hspace{5pt}$\Sigma\backslash\Sigma^{2}$\hspace{5pt} 
			& \hspace{10pt}\texttt{aa}\hspace{10pt}	& \hspace{10pt}\texttt{ab}\hspace{10pt}	& \hspace{10pt}\texttt{ba}\hspace{10pt} 	& \hspace{10pt}\texttt{bb}\hspace{10pt}	\\ \hline
\texttt{a} 	& $\lambda$								& $0$									& $0$	 									& $\lambda$			 					\\ \hline
\texttt{b}	& $\lambda$ 							& $1$									& $1$	 									& $\lambda$								\\ \hline
\end{tabular}}
\end{center}
\end{table}
\begin{table}[h]
\caption{The function $b$ after EAH$2(x)$.}
\begin{center}
\textup{
\begin{tabular}{|c|c|c|c|c|c|c|c|c|c|}
\hline
\hspace{5pt}$\Sigma\backslash\Sigma^{2}$\hspace{5pt} 
			& \hspace{10pt}\texttt{aa}\hspace{10pt}	& \hspace{10pt}\texttt{ab}\hspace{10pt}	& \hspace{10pt}\texttt{ba}\hspace{10pt} 	& \hspace{10pt}\texttt{bb}\hspace{10pt}	\\ \hline
\texttt{a} 	& $0$									& $1$									& $1$ 										& $1$		 							\\ \hline
\texttt{b}	& $1$									& $1$									& $1$ 										& $0$									\\ \hline
\end{tabular}}
\end{center}
\end{table}
\begin{table}[h]
\caption{The function $c$ after EAH$2(x)$.}
\begin{center}
\textup{
\begin{tabular}{|c|c|c|c|c|c|c|c|c|c|}
\hline
\hspace{5pt}$\Sigma\backslash\Sigma^{2}$\hspace{5pt} 
			& \hspace{10pt}\texttt{aa}\hspace{10pt}	& \hspace{10pt}\texttt{ab}\hspace{10pt}	& \hspace{10pt}\texttt{ba}\hspace{10pt} 	& \hspace{10pt}\texttt{bb}\hspace{10pt}	\\ \hline
\texttt{a} 	& $0$									& $1$									& $1$ 										& $1$									\\ \hline
\texttt{b}	& $1$									& $1$					 				& $2$ 										& $0$ 									\\ \hline
\end{tabular}}
\end{center}
\end{table}
\newline
	Let us explain these results by considering the third column of each table.
	In the second table, $b(\texttt{\textup{a}},\texttt{\textup{ba}})=1$ and
	$b(\texttt{\textup{b}},\texttt{\textup{ba}})=1$, since the substrings
	$\texttt{\textup{baa}}$ and $\texttt{\textup{bab}}$ both occur at least once in $x$.
	In the third table, $c(\texttt{\textup{a}},\texttt{\textup{ba}})=1$ is the frequency
	of $\texttt{\textup{baa}}$ in $x$, and $c(\texttt{\textup{b}},\texttt{\textup{ba}})=2$,
	since $\texttt{\textup{bab}}$ occurs twice in $x$. Thus, applying the Huffman's algorithm
	to the set of frequencies $\{1,2\}$, we encode $\texttt{\textup{a}}$ (if it is preceded by
	$\texttt{\textup{ba}}$) by $a(\texttt{\textup{a}},\texttt{\textup{ba}})=0$. Also,
	if $\texttt{\textup{b}}$ is preceded by $\texttt{\textup{ba}}$, then we encode it by
	$a(\texttt{\textup{b}},\texttt{\textup{ba}})=1$.

	Considering that the function $d$ is given by $d(1)=\texttt{\textup{aa}}$,
	$d(2)=\texttt{\textup{ab}}$, $d(3)=\texttt{\textup{ba}}$, and
	$d(4)=\texttt{\textup{bb}}$,
	one can verify that the output of \textup{EAH}$2$ in this example is the $4$-tuple:
\begin{center}
	$(\texttt{\textup{ba}},b,(0,0,1,1),01101)$,
\end{center}
	where $b$ is the function given above. Also, one can remark that the function $b$
	can be encoded using $p^{n+1}$ bits. In our example, $b$ can be encoded by $2^{3}=8$ bits,
	since $p=2$ and $n=2$.
\end{exmp}
	As one can remark, some new notations have already been used above.
	Specifically, if $\mathcal{A}$ is an algorithm and $x$ its input, then we denote by
	$\mathcal{A}(x)$ its output.
	Also, $\mathbb{N}$ denotes the set of natural numbers.
}
\\\\
{\bf Algorithm Encoder1.}
{\em
	We are now ready to describe our compression scheme based on BWT, MTF, and adaptive
	codes of order one.
	Consider the alphabet $\Sigma=\{\sigma_{0},\sigma_{1},\ldots,\sigma_{p-1}\}$ fixed above.
\textup{
\\\\
{\bf INPUT:} the string $S=s_{1}s_{2}\ldots{s_{n}}$ of length $n$ over $\Sigma$.
\begin{enumerate}
\item[1.  ]
	$\mathcal{X}:=$BWT$(S)$.
\item[2.  ]
	$\mathcal{Y}:=$MTF$(\mathcal{X}.1)$.
\item[3.  ]
	$\mathcal{Z}:=$AE$(\mathcal{Y})$.
\end{enumerate}
{\bf OUTPUT:} the $2$-tuple $(\mathcal{X}.2,\mathcal{Z})$.
}
}
\\\\
	As we have already pointed out in the beginning of this paper, our compression scheme
	performs much better on proteins than on other type of information.
	For this reason, we will report experimental results obtained only on this type of files.
	Specifically, we have tested our compressor on five well-known biological sequences:
	E.coli, hi, hs, mj, and sc. The last four files form the Protein Corpus \cite{nmw1}.
	Let us briefly describe each file separately.
\begin{description}
\item[E.coli.]
	One of the most studied biological sequences, {\em Escherichia coli} (usually abbreviated to E.coli),
	is a bacterium that lives in warm-blooded organisms.
	This genome is the only biological sequence included in the Large Canterbury Corpus \cite{ab1}.
\item[hi.]
	{\em Haemophilus influenzae} (abbreviated H.influenzae, or hi) is a bacterium that causes ear and
	respiratory infections in children.
	It was the first fully sequenced genome, made available in 1996.
	This genome is 1.83 megabases in size, and contains approximately 1740 potential genes.
	When these genes are translated into proteins, the resulting file is approximately 500 kilobytes in size
	(representing each amino acid as one byte).
\item[hs.]
	{\em Homo sapiens} (abbreviated H.sapiens, or hs) contains 5733 human genes, and the resulting
	protein file is approximately 3.3 megabytes in size.
\item[mj.]
	{\em Methanococcus jannaschii} (abbreviated M.jannaschii, or mj) lives in very hot undersea vents and
	has a unique metabolism.
	It is 1.7 megabases in size, contains 1680 genes, and the resulting protein file is approximately
	450 kilobytes in size.
\item[sc.]
	{\em Saccharomyces cerevisiae} (abbreviated S.cerevisiae, or sc) has been studied as a model organism
	for several decades. At 13 megabases in size, it is one of the largest sequenced organisms.
\end{description}
	The results reported below have been obtained by comparing Encoder1 with two of the best compressors
	available: gzip and bzip2.
\begin{description}
\item[gzip (version 1.3.3).]
	This is one of the most used UNIX utilities, and is based on Lempel-Ziv coding (LZ77).
\item[bzip2 (version 1.0.2).]
	This programme compresses files using the Burrows-Wheeler block sorting text
      compression algorithm, and Huffman coding.
	Compression is generally considerably better than that achieved by the
      LZ77/LZ78-based compressors (including gzip), and approaches the performance of the PPM
      family of statistical compressors.
\end{description}
\begin{table}[h]
\begin{center}
\textup{
Table 3. Results of compressing five protein files with gzip and bzip2.
\begin{tabular}{|l|r|r|r|r|r|r|r|}
\hline
		& Size	 	 & 			&	bits/	& 			&bits/ & \multicolumn{2}{|c|}{Improvement}	\\\cline{7-8}
File	& (bytes)	 & gzip		&	symbol	& bzip2		&symbol& bytes 		& \%		 			\\\hline
E.coli	& 4,638,690	 & 1,299,066&	2.24	& 1,251,004	& 2.16 & 48,062  	& 3.70					\\
hi		& 509,519	 & 297,517	&	4.67	& 275,412	& 4.32 & 22,105  	& 7.43					\\
hs		& 3,295,751	 & 1,897,311&	4.61	& 1,753,321	& 4.26 & 143,990 	& 7.59					\\
mj		& 448,779	 & 257,373	&	4.59	& 239,480	& 4.27 & 17,893  	& 6.95					\\
sc		& 2,900,352	 & 1,682,108&	4.64	& 1,558,813	& 4.30 & 123,295 	& 7.33					\\\hline
Total	& 11,793,091 & 5,433,375&	--		& 5,078,030	& --   & 355,345 	& -- 					\\\hline
\end{tabular}}
\end{center}
\end{table}
\begin{table}[h]
\begin{center}
\textup{
Table 4. Results of compressing five protein files with bzip2 and Encoder1.
\begin{tabular}{|l|r|r|r|r|r|r|r|}
\hline
	& Size	 & 		&	bits/	& 		&bits/ & \multicolumn{2}{|c|}{Improvement}\\\cline{7-8}
File	& (bytes)	 & bzip2	&	symbol& Encoder1	&symbol& bytes	& \%				\\\hline
E.coli& 4,638,690	 & 1,251,004&	2.16	& 1,159,813	& 2.00 & 91,191	& 7.29			\\
hi	& 509,519	 & 275,412	&	4.32	& 274,115	& 4.30 & 1,297	& 0.47			\\
hs	& 3,295,751	 & 1,753,321&	4.26	& 1,728,061	& 4.19 & 25,260	& 1.44			\\
mj	& 448,779	 & 239,480	&	4.27	& 238,294	& 4.25 & 1,186	& 0.50			\\
sc	& 2,900,352	 & 1,558,813&	4.30	& 1,539,390	& 4.25 & 19,423	& 1.25			\\\hline
Total	& 11,793,091 & 5,078,030&	--	& 4,939,673	& --	 & 138,357	& --				\\\hline
\end{tabular}}
\end{center}
\end{table}
	Given the results reported here, one can conclude that our compression scheme is one of the
	most competitive algorithms in the field of biological data compression.

	Further work in this field is intended to compare our compression scheme with
	some of the best PPM techniques as they are being developed for (biological) data compression.
	We welcome any suggestions or comments, especially from the readers interested in these matters.



\end{document}